\documentclass[prl,twocolumn,amsmath,amssymb]{revtex4}
\usepackage{graphicx}
\usepackage{dcolumn}
\usepackage{bm}
\usepackage{color}
\usepackage{hyperref} 

\begin{document}

\title{Shocks in nonlocal media}
\author{Neda Ghofraniha,$^{1}$ Claudio Conti,$^{2,3}$ Giancarlo Ruocco,$^{3,4}$ Stefano Trillo$^{5}$}
\email{claudio.conti@phys.uniroma1.it}
\affiliation{
$^{1}$ Research Center SMC INFM-CNR, Universit\`{a} di Roma ``La Sapienza'',  
P. A. Moro 2, 00185, Roma, Italy\\
$^{2}$Centro Studi e Ricerche ``Enrico Fermi'', Via Panisperna 89/A, 
00184 Rome, Italy \\
$^{3}$Research center SOFT INFM-CNR Universit\`{a} di Roma ``La Sapienza'',  P. A. Moro 
2, 00185, Roma, Italy\\
$^{4}$Dipartimento di Fisica, Universit\`{a} di Roma ``La Sapienza'',  P. A. Moro 
2, 00185, Roma, Italy\\
$^{5}$ Dipartimento di Ingegneria, Universit\`{a} di Ferrara, Via Saragat 1, 
44100 Ferrara, Italy
}
\date{\today}  
\begin{abstract}
We investigate the formation of collisionless shocks along the spatial profile
of a gaussian laser beam propagating in nonlocal nonlinear media. For defocusing
nonlinearity the shock survives the smoothing effect of the nonlocal response, though its
dynamics is qualitatively affected by the latter, whereas for focusing nonlinearity it dominates over filamentation. 
The patterns observed in a thermal defocusing medium
are interpreted in the framework of our theory.
\end{abstract}

\pacs{42.65.Jx, 42.65.Tg, 82.70.-y}	

\maketitle
Shock waves are a general phenomenon thoroughly investigated in 
disparate area of physics (fluids and water waves, 
plasma physics, gas dynamics, sound propagation,  physics of explosions, etc.),
entailing the propagation of discontinuous solutions typical of hyperbolic PDE models
\cite{Whitman74,Liberman86}. 
They are also expected in (non-hyperbolic) universal models for dispersive nonlinear media,
such as the Korteweg-De Vries (KdV) and nonlinear Schr\"{o}dinger 
(NLS, or analogous Gross-Pitaevskii) equations, 
since hydrodynamical approximations of such models hold true in certain regimes 
(typically, in the weakly dispersive or strongly nonlinear case) \cite{gp,wkb,Kamchatnov02}. 
However, in the latter  cases, no true discontinuous solutions are permitted.  
The general scenario, first investigated  by Gurevich and Pitaevskii \cite{gp}, 
is that dispersion regularizes the shock, determining the onset of oscillations
that appear near wave-breaking points and expand afterwards.
This so-called collisionless shock has been observed for example in ion-acoustic waves
\cite{Taylor70}, or wave-breaking of optical pulses 
in a normally dispersive fiber \cite{Grischkowsky89}, 
and recently in a Bose-Einstein condensate with positive scattering length \cite{Perez-Garcia04}.\\
\indent
In this Letter we investigate how nonlocality of the nonlinear response
affects the formation of a collisionless shock in a system ruled by a NLS model. 
In fact nonlocality plays a key role in many physical systems due to
transport phenomena and finite range interactions (e.g. as in Bose-Einstein condensation), 
and can be naively thought to smooth and eventually wipe out steep fronts characteristic of shocks.
More specifically, we place this problem in the context of nonlinear optics
where nonlocality arises quite naturally in different media \cite{Wyller02, Conti, Yakimenko05, Segev05}, 
studying the spatial propagation of  a fundamental (gaussian TEM$_{00}$) laser mode subject
to diffraction and nonlocal focusing/defocusing action (Kerr effect).
In a {\em  defocusing} and ideal (local and lossless) medium, 
high intensity portions of the beam diffract more rapidly than the tails 
leading, at sufficiently high powers, to overtaking and oscillatory wave-breaking
similar (in 1D) to what observed in the temporal case
\footnote{
paraxial diffraction in defocusing media is well known to be isomorphus in 1D
to propagation in a normally dispersive focusing medium  
as considered in Ref. \cite{Grischkowsky89}
}.
We find that, while shock is not hampered by the presence of 
(even strong) nonlocality, the mechanism of its formation as well as
post-shock patterns are qualitatively affected by the nonlocality.
Experimental results obtained with a thermal defocusing nonlinearity are consistent
with our theory and shed new light on the interpretation of the thermal lensing phenomenon.\\
\indent Importantly, our theory permits also to establish that nonlocality allows the shock to form also
in the {\em focusing} regime where, contrary to the local case, it can prevails over filamentation 
or modulational instability (MI).

\begin{figure}
\includegraphics[width=8.3cm]{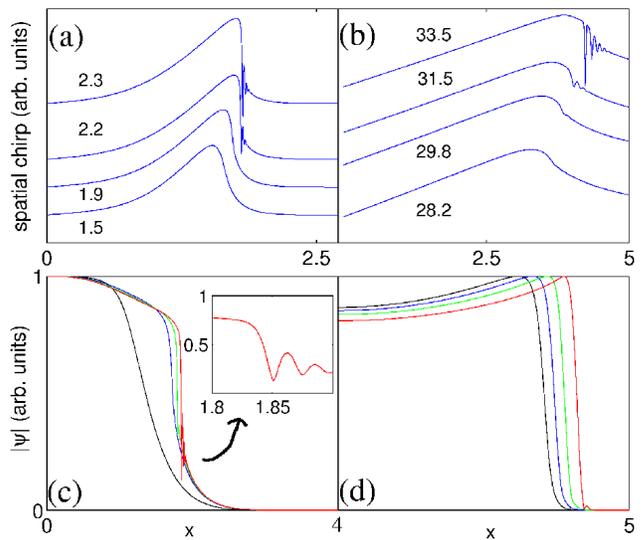}
\caption{ (Color online) 1D spatial profiles of phase chirp  $u(x)$ (a-b) 
and amplitude $|\psi(x)|=\sqrt{\rho(x)}$ (c-d),
as obtained from Eqs. (\ref{nls1}-\ref{nls2}) with $\varepsilon =10^{-3}$,
$\alpha=0$, $\chi=-1$ (defocusing), and  $\psi_{0}=\exp(-x^2)$, for different $z$ as indicated:
(a-c) local case, $\sigma^2=0$; (b-d) nonlocal case, $\sigma^2=5$.
\label{figure1}}
\end{figure}

{\it Theory} We start from the paraxial wave equation obeyed by
the envelope $A$ 
of a monochromatic field $E=(\frac{2}{c\epsilon_0 n})^{1/2}A \exp(i k Z -i \omega T)$  ($|A|^{2}$ is the intensity)
\begin{equation} \label{paraxial}
i  \frac{\partial A}{\partial Z} +\frac{1}{2k} \left( \frac{\partial^2 A}{\partial X^2} + \frac{\partial^2 A}{\partial Y^2} \right) 
+ k_0 \Delta n A =- i \frac{\alpha_0}{2} A \text{.}
\end{equation}
where $k=k_{0} n=\frac{\omega}{c} n$ is the wave-number, and $\alpha_{0}$ the intensity loss rate.
A sufficiently general nonlocal model can be obtained by coupling Eq. (1)
to an equation that rules the refractive index change  $\Delta n$ of nonlinear origin.
Introducing the scaled coordinates $x,y,z=X/w_0,Y/w_0,Z/L$, 
and complex variables $\psi=A/\sqrt{I_0}$ and $\theta=k_{0} L_{nl} \Delta n$,
where $L_{nl}=(k_{0} |n_{2}| I_0)^{-1}$ is the nonlinear length scale associated with peak intensity $I_{0}$
and a local Kerr coefficient $n_{2}$ ($\Delta n=n_{2} |A|^{2}$), $L_d= k  w_0^2$ is the characteristic diffraction 
length associated with the input spot-size $w_{0}$,
and $L \equiv \sqrt{L_{nl} L_d}$,  such model can be conveniently written as follows  \cite{Conti}
\begin{eqnarray}
\displaystyle i  \varepsilon \frac{\partial \psi}{\partial z} +
\frac{ \varepsilon^2}{2} \nabla^2_{\perp}
\psi + \chi \theta \psi=-i \frac{\alpha}{2} \varepsilon \psi, \label{nls1} \\
-\sigma^2 \nabla^2_{\perp} \theta + \theta = |\psi|^2, \label{nls2}
\end{eqnarray}
where $\alpha=\alpha_{0} L$, $\nabla^{2}_{\perp}=\partial_{x}^{2} + \partial_{y}^{2}$,
$\chi=n_{2}/|n_{2}| =\pm 1$ is the sign of the nonlinearity,
and $\sigma^{2}$ is a free parameter that measures the {\em degree of nonlocality}.
The peculiar dimensionless form of Eqs. (\ref{nls1}-\ref{nls2}) 
where $\varepsilon \equiv L_{nl}/L=\sqrt{L_{nl}/L_d}$ is a small quantity,
highlights the fact that we will deal with the weakly diffracting (or strongly nonlinear) regime,
such that the local $\sigma=0$ and lossless $\alpha=0$ limit 
yields a semiclassical Schr\"odinger equation with cubic potential
($\varepsilon$ and $z$ replace Planck constant and time, respectively).
We study Eqs. (\ref{nls1}-\ref{nls2}) subject to the axi-symmetric  
gaussian input $\psi_{0}(r)=\exp(-r^2)$, $r \equiv \sqrt{x^2+y^2}$,
describing a fundamental laser mode at its waist.
For $\varepsilon \ll 1$, its evolution can be studied in the framework of the 
WKB trasformation $\psi(r,z)=\sqrt{\rho(r,z)} \exp \left[i \phi(r,z) \right/\varepsilon]$ \cite{wkb}.
Substituting in Eqs. (\ref{nls1}-\ref{nls2}) and retaining only leading orders in $\varepsilon$, we obtain
\begin{eqnarray}
\rho_{z} + \left[ \frac{(D-1)}{r} \rho u + (\rho u)_{r} \right] &=& - \alpha \rho;\;\;
u_{z} + u u_{r} - \chi \theta_{r}=0,\nonumber \\
 -\sigma^{2} \left( \theta_{rr} + \frac{D-1}{r} \theta_{r} \right) + \theta &=& \rho.
\label{sweqs}
\end{eqnarray}
where $u\equiv \phi_r$ is the phase chirp, and $D=2$ is the transverse dimensionality. 
The 1D case described by Eqs. (\ref{sweqs}) with $D=1$ and $r \rightarrow x$ ($\partial_{y}=0$)
illustrates the basic physics with least complexity. In the defocusing case ($\chi=-1$)
for an ideal medium ($\sigma=\alpha=0$, $\theta=\rho$), Eqs. (\ref{sweqs}) are a well known
hyperbolic system of conservation laws (Eulero and continuity equations) with real
celerities (or eigenspeeds, i.e. velocities of Riemann invariants) $v^{\pm}=u \pm \sqrt{-\chi \rho}$,
which rules gas dynamics ($u$ and $\rho$ are velocity and mass density of a gas with pressure $\propto \rho^2$).
A gaussian input  is known to develop two symmetric shocks at finite $z$ \cite{wkb}.  
Importantly the diffraction, which is initially of order $\varepsilon^2$, 
starts to play a major role in the proximity of the overtaking point, 
and regularize the wave-breaking through the appearance of 
fast (wavelength $\sim \varepsilon$) oscillations which connect
the high and low sides of the front and expand outwards (far from the beam center) \cite{gp}. 
Such oscillations, characteristic of a collisionless shock, appear simultaneously 
in intensity and phase chirp ($u$) as clearly shown in Fig. \ref{figure1}(a,c).

\begin{figure}
\includegraphics[width=8.3cm]{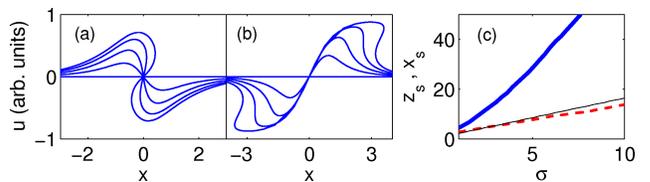}
\caption{(Color online)  (a) $u(x)$ for different $z$ and $\chi=1$ (focusing), $\sigma=1$;
(b) as in (a) for $\chi=-1$ (defocusing); (c) shock distance $z_{s}$ ($\chi= -1$ bold solid,  $\chi= 1$ thin solid) 
and shock position $x_{s}$ in the defocusing case (dashed line).
\label{figODE}} 
\end{figure}

In the nonlocal case, the index change $\theta(x)$ can be wider than the gaussian mode
(for large $\sigma$) and the shock dynamics is essentially driven by the chirp $u$ 
with $\rho$ adiabatically following. This can be seen by means of the
following approximate solution of Eqs. (\ref{sweqs}):
considering that the equation for $\rho$ is of lesser order [$O(\epsilon)$],
with respect to those for $\theta$ and $u$ [$O(1)$], we assume $\rho=\exp(-2x^{2})$ unchanged
in $z$ and solve exactly the third of Eq. (\ref{sweqs}) for $\theta(x)$ (though derived easily, its full expression is quite cumbersome). 
Then, applying the theory of characteristics \cite{Whitman74}, 
the second of Eqs. (\ref{sweqs}) is reduced to the following ODEs,
where dot stands for $d/dz$
\begin{equation}
\dot{x}=u\;;\; \dot{u}=\chi \theta_x,
\label{ode}
\end{equation}
equivalent to the motion of a unit mass in the potential $V(x)=-\chi \theta$
with conserved energy $E=\frac{u(z)^2}{2}+V(x)$. 
The solution of Eqs. (\ref{ode}) with initial condition $x(0)=s, u(0)=0$ yields
$x(s,z), u(s,z)$ in parametric form, from which overtaking is found whenever
$u(x,z)$ (obtained by eliminating $s$) becomes a multivalued function of $x$ at finite $z=z_{s}$. 
The shock point  corresponding to $|du/dx| \rightarrow \infty$ is found from
the solution $u(x,z)$ displayed in Fig. \ref{figODE}(a) [ \ref{figODE}(b)], 
at positions $x=\pm x_{s} \neq 0$ (defocusing case) or $x_{s}=0$
(focusing case). The shock distance $z_{s}$ increases with $\sigma$ in both cases,
as shown in Fig. \ref{figODE}(c). 

We have tested these predictions by integrating numerically Eqs. (\ref{nls1}-\ref{nls2}). 
Simulations with $\chi=-1$ [see Fig. \ref{figure1}(b,d)] show indeed steepening
and post-shock oscillations in the spatial chirp $u$, which are
accompanied by a steep front in $\rho$ moving outward.
The shock location in $x$ and $z$ is in good agreement with the results 
of our approximate analysis summarized in Fig. \ref{figODE}.

\begin{figure}
\includegraphics[width=8.3cm]{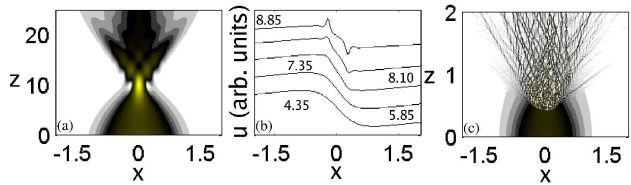} 
\caption{ (Color online) Level plot of the intensity in the
focusing case ($\chi=1$, $\varepsilon=0.01$): 
(a) nonlocal case ($\sigma^{2}=25$); 
(b) chirp profile for various $z$ for (a);
(c) quasi-local case ($\sigma^{2}=10^{-5}$).
\label{focusing}} 
\end{figure}

Numerical simulations of  Eqs. (\ref{nls1}-\ref{nls2}) validates also 
the focusing scenario. The field evolution displayed in Fig. \ref{focusing}(a)
exhibits shock formation at the focus point ($x_{s}=0, z_{s} \simeq 8$, for $\sigma=5$) 
driven the phase whose chirp is shown in Fig. \ref{focusing}(b). This is remarkable because, {\em in the local limit} $\sigma=0$, 
the celerities become imaginary (the equivalent gas would have pressure decreasing with increasing density $\rho$),
and no shock could be claimed to exist. In this limit, the reduced problem (\ref{sweqs}) is elliptic
and the initial value problem is ill-posed \cite{focusing}, an ultimate consequence of the
onset of MI: modes with transverse (normalized) wavenumber $q<\Delta q$
grow exponentially with $z$, with both gain $g$ and bandwidth $\Delta q$ scaling as $1/\varepsilon$. 
However, the nonlocal response tends to frustrate MI (see also Refs. \cite{Wyller02,Conti}),
as shown by standard linear stability analysis which yields $g=\sqrt{d(2\overline\chi -d)/\varepsilon^2}$
(we set $d \equiv \varepsilon^2 q^2/2$  and $\overline\chi \equiv \chi/(1+\sigma^2 q^2)$),
in turn implying a strong reduction of both gain and bandwidth for large $\sigma$.
In order to emphasize the difference between the local and nonlocal regime, 
we contrast Fig. \ref{focusing}(a) with the analogous evolution
[see Fig. \ref{focusing}(c)] obtained in the quasi-local limit ($\sigma^2=10^{-5}$), 
which appears to be clearly dominated by filamentation. 

{\em Thermal nonlinearity} 
The physics of the defocusing case can be experimentally tested 
by exploiting thermal nonlinearities of strongly absorptive bulk samples,
that we show below to fit our model. In this case, the system
relaxes to a steady-state refractive index change $\Delta n=(dn/dT) \Delta T$, 
where $dn/dT$ is the thermal coefficient, and $\Delta T$ 
the local temperature change due to optical absorption.
It is well known that this so-called thermal lens distorts a laser beam propagating in the medium
\cite{tl,rings,Brochard97}. However, only perturbative approaches to the problem have been proposed 
(ray optics or Fresnel diffraction theory is applied after the lens profile is 
worked out from gaussian ansatz \cite{tl}),
while the role of shock phenomena was completely overlooked.
\begin{figure}[h]
\includegraphics[width=8.3cm]{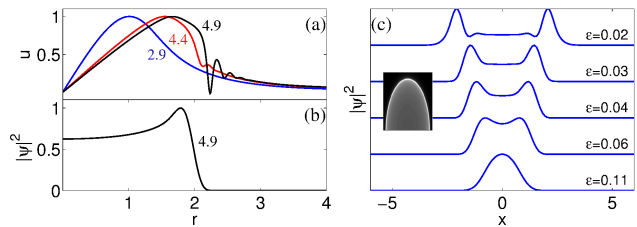}
\caption{2D evolution according to Eqs. (\ref{nls1}-\ref{nls2}) with $\sigma^2= 1$, $\alpha=1$:
(a) radial phase chirp at different $z$, as indicated, showing steepening
and shock formation for $\varepsilon=10^{-2}$;
(b) corresponding intensity profile $|\psi|^2$ (maximum scaled to unity) at $z=4.9$;
(c) transverse intensity profile vs. $x$ (at $y=0$) at $z=1/(4 \varepsilon)$
and different values of $\varepsilon$ ($\alpha_0=62 $cm$^{-1}$, $\sigma=0.3$).
} 
\label{fig2D} \end{figure}

We assume that the temperature field $\Delta T=\Delta T_{\perp}(X,Y)$
obeys the following 2D heat equation 
\begin{equation} \label{thermaleq1}
(\partial_{X}^2 + \partial_{Y}^2) \Delta T_\perp - C \Delta T_\perp =-\gamma |A|^{2}
\end{equation}
where the source term account for absorption proportional
to intensity through the coefficient $\gamma=\alpha_0/(\rho_0 c_p D_{T})$,  where $\rho_0$ the material density, 
$c_p$ the specific heat at constant pressure, and $D_T$ is the thermal diffusivity (see e.g. \cite{Brochard97}). 
Eq. (\ref{thermaleq1}) has been already employed to model a refractive index of  thermal origin
in Ref. \cite{Yakimenko05}, and in Ref. \cite{Segev05} in the limit $C=0$ which
is equivalent to consider the range of nonlocality (measured by $1/C$, see below) to be infinite.
Starting from the 3D heat equation $\nabla^2 \Delta T=-\gamma |A|^{2}$,
the latter regime amounts to assume $\Delta T(X,Y,Z)=\Delta T_\perp(X,Y)$,
which is justified when longitudinal changes in intensity $|A|^{2}$ are negligible
as for solitary (invariant in $Z$) wave-packets in the presence of low absorption \cite{Segev05}.
Viceversa, in the regime of strong absorption, we need to account for longitudinal
temperature profiles that are known from solutions of the 3D heat equations  
to be peaked at characteristic distance $\hat Z$ in the middle of sample 
and decay to room temperature on the facets \cite{tl}. Since highly nonlinear phenomena
occurs in the neighborhood of $\hat Z$ where the index change is maximum,
we can use a (longitudinal) parabolic approximation with characteristic width $L_{eff} (\sim L$) 
of the 3D temperature field $\Delta T(X,Y,Z)= \left[1- \frac{(Z-\hat{Z})^2}{ 2 L_{eff}^2} \right] \Delta T_\perp(X,Y)$
and consequently approximate $\nabla^2 \Delta T \simeq (\partial_{X}^2 + \partial_{Y}^2) \Delta T_\perp - L_{eff}^{-2} \Delta T_\perp$, 
so that the 3D heat equation reduces to Eq. (\ref{thermaleq1}) with $C=1/L_{eff}^2$.
Following this approach, Eq. (\ref{thermaleq1}) coupled to Eq. (\ref{paraxial}) 
can be casted in the form of Eqs. (\ref{nls1}-\ref{nls2}) by posing
$\theta=k_{0} L_{nl} |dn/dT| \Delta T_{\perp}$ and $\sigma^{2}=1/(Cw_0^2)=L_{eff}^2/w_0^2$.
The model reproduces the infinite range nonlocality for negligible losses 
($L_{eff} \rightarrow \infty$); while for thin samples [$|(\partial_{X}^2 + \partial_{Y}^2) \Delta T_\perp|<< |L_{eff}^{-2} \Delta T_\perp|$],
$L_{eff}$ can be related to the Kerr coefficient $n_{2}$ as
\begin{equation}
L_{eff}=\sqrt{\frac{ |n_{2}|}{\gamma |dn/dT|}}=\sqrt{\frac{D_T \rho_0 c_p |n_2|}{\alpha_0 |dn/dT|}} 
\label{n2leff}
\end{equation}
which establishes a link between the degree of nonlocality and the strength of the nonlinear response 
(similarly to other nonlocal materials \cite{Conti}).\\
\indent 
Having retrieved the model Eqs. (\ref{nls1}-\ref{nls2}), let us show next that the scenario illustrated previously 
applies substantially unchanged in bulk (2D case) even on account for the optical power loss
($\alpha \neq 0$). An example of the general dynamics is shown in Fig. \ref{fig2D},
where we report a simulation of the full model (\ref{nls1}-\ref{nls2}),
with $\sigma^{2}=1$ and relatively large loss $\alpha=1$.
In analogy to the 1D case, Fig. \ref{fig2D}(a) clearly shows that the radial chirp $u=\phi_{r}$ steepens and then develop
characteristic oscillations after the shock point ($z \simeq 6$, where $|\partial_{r} u| \rightarrow \infty$). 
Correspondingly the intensity exhibits also an external front which is connected to a flat central region
with a characteristic overshoot [see Fig. \ref{fig2D}(b)] corresponding to a brighter ring [inset in  Fig. \ref{fig2D}(c)]. 
For larger distances this structure moves outward following the motion of the shock.
In the experiment such motion can be observed, at fixed physical lenght, by increasing
the power, which amounts to decrease $\varepsilon$ while scaling $z$ and $\alpha$ accordingly
($z\propto 1/\varepsilon$, $\alpha\propto\varepsilon$), as displayed in Fig. \ref{fig2D}(c) for $\sigma=0.3$.

\begin{figure}
\includegraphics[width=8.3cm]{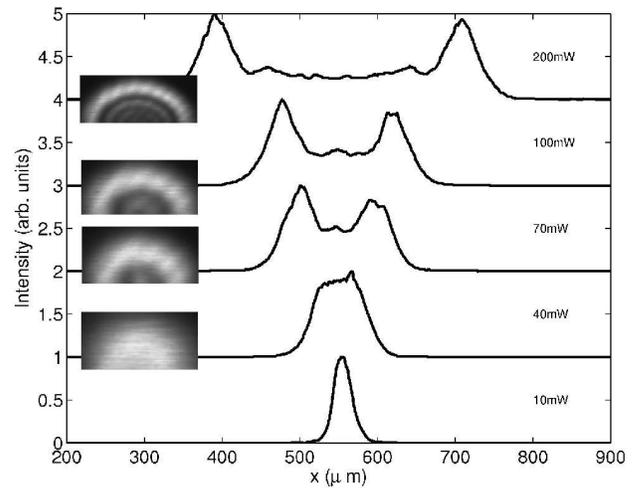}
\caption{Radial profiles of intensity observed in the 
thermal medium for different input powers.
The insets show the corresponding 2D output patterns.
\label{exp1}} 
\end{figure}

As a sample of a strongly absorbing medium we choose a $1$ mm long cell  
filled with an acqueous solution of Rhodamine B ($0.6$ mM concentration).
Our measurements of the linear and nonlinear properties of the sample 
performed by means of the Z-scan technique gives data consistent with
the literature \cite{Sinha00}, and allows us to extrapolate at the operating vacuum wavelength of 532 nm,
a linear index $n=1.3$, a defocusing nonlinear index $n_{2}=7 \times 10^{-7}$ cm$^2$W$^{-1}$,
and $\alpha_{0}=62$ cm$^{-1}$. 
For our sample $D_T=1.5\times 10^{-7}$ m$^2$s$^{-1}$, $\rho_0=10^3$ kg m$^{-3}$, $c_p=4\times 10^3 J kg^{-1} K^{-1}$ and $|dn/dT|=10^{-4}$ K$^{-1}$ 
($\gamma \cong 10^4$ K W$^{-1}$), and exploiting Eq.  (\ref{n2leff}) we estimate 
$L_{eff} \cong 10 \mu$m 
($L_{eff}<<L$ because of the strong absorption that causes strong heating of our sample near the input facet), 
and correspondingly the degree of nonlocality $\sigma \cong 0.3$.
We operate with an input gaussian beam with fixed intensity waist
$w_{0I}=w_0/\sqrt{2}=20~\mu$m ($L_{d} \cong 12$ mm) focused onto the input face of the cell.
With these numbers, an input power $P=\pi w_{0I}^{2} I_{0}=200$ mW yields a nonlinear
length $L_{nl} \cong 8~\mu$m ($L \cong 0.3$ mm), 
which allows us to work in the semiclassical regime with $\varepsilon \cong 0.025$. 
The radial intensity profiles together with the 2D patterns imaged by means of a $40\times$ 
microscope objective and a recording CCD camera are reported  in Fig. \ref{exp1}.
As shown the beam exhibits the formation of the bright ring whose external
front moves outward with increasing power, consistently with the reported simulations. 
We point out that, at higher powers, we observe (both experimentally and numerically) 
that the moving intensity front leaves behind damped oscillations that correspond
to inner rings of lesser brightness, as reported in literature \cite{rings}. This, however,
occurs well beyond the shock point that we have characterized so far.
\newline \indent
In summary,  the evolution of a gaussian beam
in the strong nonlinear regime is characterized by occurence
of collisionless (i.e., regularized by diffraction) shocks that 
survive the smoothing effect of (even strong) nonlocality.
While experimental results support the theoretical scenario 
in the defocusing case, the remarkable result that the nonlocality
favours shock dynamics over filamentation requires future investigation.

\end{document}